\documentclass[conference]{IEEEtran} 
\IEEEoverridecommandlockouts
\usepackage{cite}
\usepackage{amsmath,amssymb,amsfonts}
\usepackage{algorithmic}
\usepackage{graphicx}
\usepackage{textcomp}
\usepackage{xcolor}
\usepackage{ textcomp }
\def\BibTeX{{\rm B\kern-.05em{\sc i\kern-.025em b}\kern-.08em
    T\kern-.1667em\lower.7ex\hbox{E}\kern-.125emX}}
\begin{document}

\title{Temporal Correlation of Internet \\ Observatories and Outposts
\thanks{This material is based upon work supported by the Assistant Secretary of Defense for Research and Engineering under Air Force Contract No. FA8702-15-D-0001, National Science Foundation CCF-1533644, and United States Air Force Research Laboratory and Air Force Artificial Intelligence Accelerator Cooperative Agreement Number FA8750-19-2-1000. Any opinions, findings, conclusions or recommendations expressed in this material are those of the author(s) and do not necessarily reflect the views of the Assistant Secretary of Defense for Research and Engineering, the National Science Foundation, or the United States Air Force. The U.S. Government is authorized to reproduce and distribute reprints for Government purposes notwithstanding any copyright notation herein.}
}

\author{\IEEEauthorblockN{Jeremy Kepner$^1$, Michael Jones$^1$, Daniel Andersen$^2$, Ayd{\i}n Bulu{\c{c}}$^3$, Chansup Byun$^1$,   K Claffy$^2$, Timothy Davis$^4$,  \\ William Arcand$^1$, Jonathan Bernays$^1$, David Bestor$^1$, William Bergeron$^1$, Vijay Gadepally$^1$,  \\ Daniel Grant$^5$, Micheal Houle$^1$, Matthew Hubbell$^1$,  Hayden Jananthan$^1$, Anna Klein$^1$, Chad Meiners$^1$, \\ Lauren Milechin$^1$, Andrew Morris$^5$, Julie Mullen$^1$, Sandeep Pisharody$^1$, Andrew Prout$^1$,  Albert Reuther$^1$, \\ Antonio Rosa$^1$, Siddharth Samsi$^1$, Doug Stetson$^1$, Charles Yee$^1$, Peter Michaleas$^1$
\\
\IEEEauthorblockA{$^1$MIT,  $^2$CAIDA, $^3$LBNL, $^4$Texas A\&M, $^5$GreyNoise
}}}
\maketitle

\begin{abstract}
The Internet has become a critical component of modern civilization requiring scientific exploration akin to endeavors to understand the land, sea, air, and space environments. Understanding the baseline statistical distributions of traffic  are essential to the scientific understanding of the Internet.  Correlating data from different Internet observatories and outposts can be a useful tool for gaining insights into these distributions.   This work compares observed sources from the largest Internet telescope (the CAIDA darknet telescope) with those from a commercial outpost (the GreyNoise honeyfarm).  Neither of these locations actively emit Internet traffic and provide distinct observations of unsolicited Internet traffic (primarily botnets and scanners).  Newly developed GraphBLAS hyperspace matrices and D4M associative array technologies enable the efficient analysis of these data on significant scales.  The  CAIDA sources are well approximated by a Zipf-Mandelbrot distribution.  Over a 6-month period 70\% of the brightest (highest frequency) sources in the CAIDA telescope are consistently detected by coeval observations in the GreyNoise honeyfarm.  This overlap drops as the sources dim (reduce frequency) and as the time difference between the observations grows.  The probability of seeing a CAIDA source is proportional to the logarithm of the brightness.  The temporal correlations are well described by a modified Cauchy distribution.  These observations are consistent with a correlated high frequency beam of sources that drifts on a time scale of a month.
  
\end{abstract}


\begin{figure}
\center{\includegraphics[width=0.65\columnwidth]{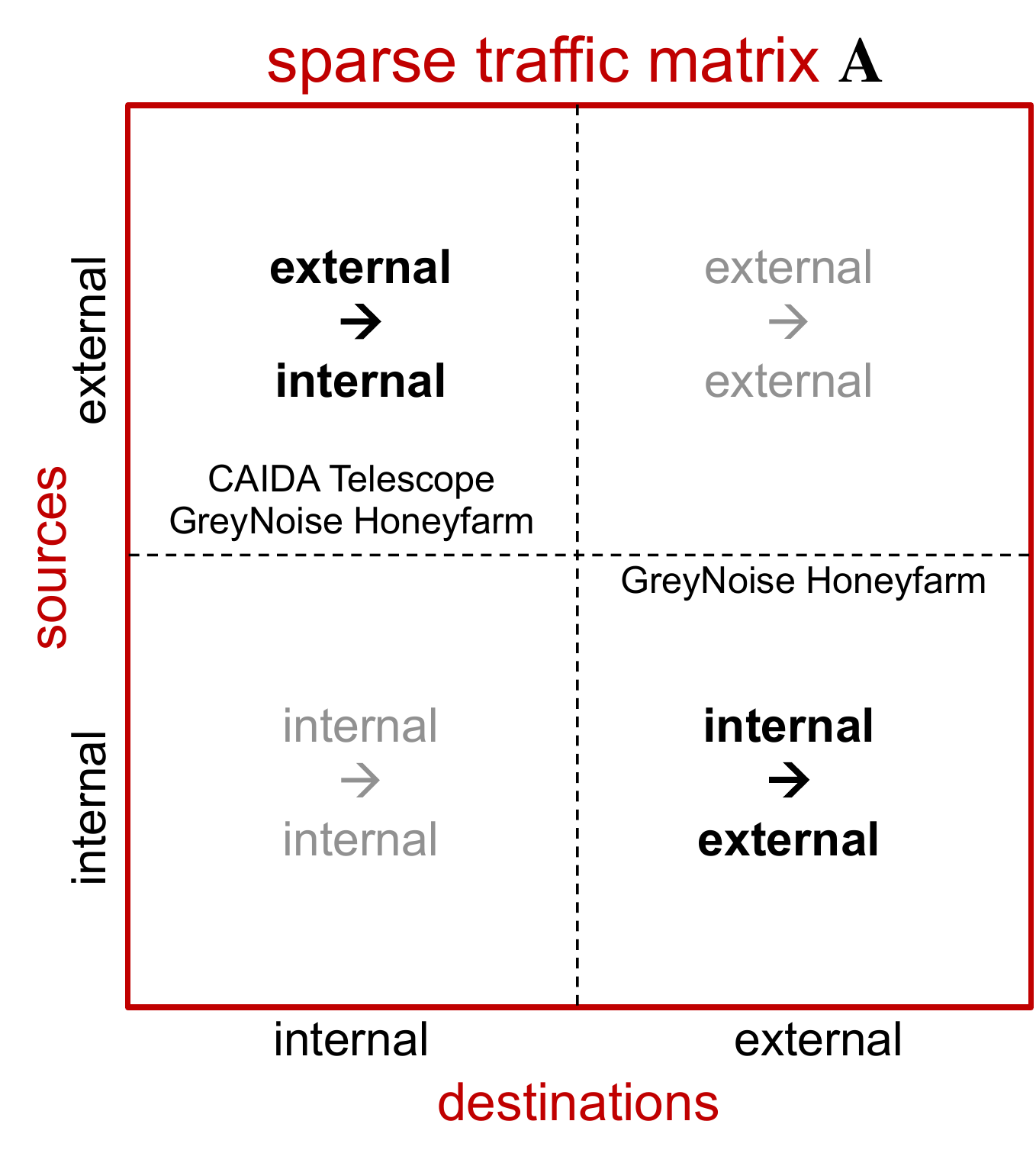}}
      	\caption{{\bf Network Traffic Matrix.} The traffic matrix of an Internet observatory or outpost  can be divided into quadrants separating internal and external traffic.  The CAIDA Telescope Internet observatory monitors a darkspace, so only the upper left (external $\rightarrow$ internal) quadrant will have data.  The GreyNoise honeyfarm outpost deduces more information about sources by responding to traffic to determine its nature and so exists in both the upper left (external $\rightarrow$ internal) quadrant and the lower right (internal  $\rightarrow$ external) quadrant.} 
      	\label{fig:GatewayTrafficMatrix}
\end{figure}

  The Internet has become as essential as land, sea, air, and space for enabling activities as diverse as commerce, education, health, and entertainment \cite{Cisco2018-2023}.   Understanding the Internet is likewise as important as studying these other domains \cite{kepner2021zero}.  Developing scientific insights on how the Internet behaves requires observations and data \cite{claffy2000measuring, li2013survey, rabinovich2016measuring, ClaffyClark2020}. In the cyber domain,  observatories and outposts have been constructed to gather data on Internet traffic and provide a starting point for scientific exploration of the Internet \cite{CAIDA2019, CAIDA2022, GCA2022, Greynoise2022, MAWI2020, Shadowserver2022, kepner2020multi}.  Correlating data from different Internet observatories and outposts can be a useful tool for gaining insights into these questions.  Specifically, comparing observations of the Internet from two different viewpoints at the same time can tell us which measurements are consistent.

Internet measurement dates to its inception.  The need for better Internet measurement has grown with the importance of the Internet and the corresponding increase in cyber threats.  Recent Internet measurements shed light on a range of important questions, such as the connectivity of public clouds \cite{marder2021inferring}, Geolocating Internet routers \cite{luckie2021learning}, identifying state-owned Internet operators \cite{carisimo2021identifying}, the statistical shape of video viewership \cite{manousis2021shape}, mapping the domains of e-mail service providers \cite{liu2021s}, detecting Internet-scale surveillance devices \cite{yan2021detecting}, and botnet detection \cite{cheng2021botnet}.  Vantage point plays a key role in these measurements hightlighting the need for correlating measurements.  For example, recent work on distributed denial of service (DDoS) attacks \cite{nawrocki2021far} indicates that ``Surprisingly, IXPs [Internet eXchange Points] and honeypots observe mostly disjoint sets of attacks: 96\% of IXP-inferred attacks were invisible to a sizable honeypot platform.'' 

  The largest public Internet observatory is the Center for Applied Internet Data Analysis (CAIDA) Telescope that operates a variety of sensors including a continuous stream of packets from an unsolicited darkspace representing approximately 1/256 of the Internet.  CAIDA Telescope data consists of almost entirely malicious traffic.  The scale and continuous duration of the CAIDA Telescope makes it ideally suited for correlation studies with smaller outposts like the GreyNoise honeyfarm that engage more deeply with Internet sources (see Figure~\ref{fig:GatewayTrafficMatrix}).  Using selected contiguous samples of $2^{30}$ CAIDA packets and comparing the GreyNoise database over a 15 month period allows analysis of the spatial and temporal patterns of Internet traffic and directly addresses the question of what is similar and what is different  as time and location changes.

The outline of the rest of the paper is as follows.  First, the CAIDA Telescope and GreyNoise honeyfarm data sets are described.  Second, the relevant network quantities and their distributions are defined in terms of traffic matrices.   Third, the CAIDA-GreyNoise correlations are presented along with various empirical fits to the data followed by a discussion of these results.  Finally, our conclusions and directions for further work are presented.

\section{Data Sets}

  The CAIDA Telescope monitors an Internet darkspace (also referred to as a black hole, Internet sink, or darknet) that is a globally routed /8 network that carries almost no legitimate traffic because there are few allocated addresses in this Internet prefix. After discarding the small amount of legitimate traffic from the incoming packets, the remaining data represent a continuous view of anomalous unsolicited traffic, or Internet background radiation. Almost every computer on the Internet will receive some form of this background traffic.  This unsolicited traffic results from a wide range of events, such as backscatter from randomly spoofed sources used in denial-of-service attacks, the automated spread of Internet worms and viruses, scanning of address space by attackers or malware looking for vulnerable targets, and various misconfigurations (e.g. mistyping an IP address). In recent years, traffic destined to darkspace has evolved to include longer-duration, low-intensity events intended to establish and maintain botnets.  CAIDA personnel maintain and expand the telescope instrumentation, collecting, curating, archiving, and analyzing the data to enable data access for vetted researchers.

\begin{table}[htp]
\caption{GreyNoise and CAIDA Data Sets.}
\vspace{-0.25cm}
Data collection start time, collection duration, and number of unique sources from the GreyNoise and CAIDA data sets.  GreyNoise data was collected for each month. The sharp increases in 2020-03 and 2021-04 are a result of configuration changes.  $2^{30}$ packets of CAIDA data were selected approximately every 6 weeks on Wednesdays either at noon or midnight.  Constant packet, variable time samples simplify the statistical analysis of the heavy-tail distributions commonly found in network traffic quantities \cite{kepner19hypersparse, nair2020fundamentals, kepner2022new}. \center{\includegraphics[width=1.0\columnwidth]{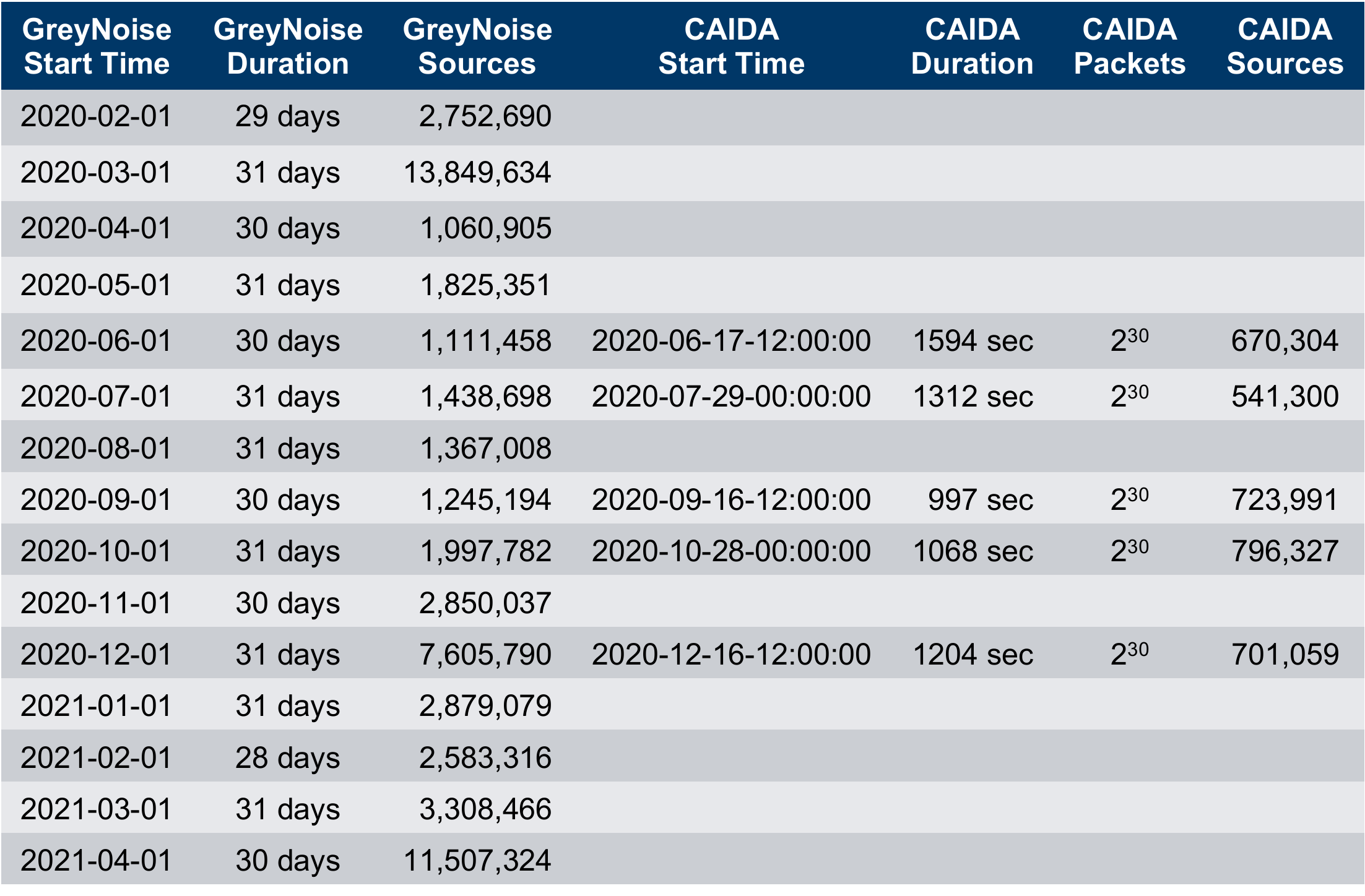}}
\label{tab:GreyNoise-CAIDA-Data}
\end{table}%

The CAIDA Telescope monitors the traffic into and out of a set of network addresses providing a natural observation point of network traffic.  These data can be viewed as a traffic matrix where each row is a source and each column is a destination.  The CAIDA Telescope traffic matrix can be partitioned into four quadrants (see Figure~\ref{fig:GatewayTrafficMatrix}).  These quadrants represent different flows between nodes internal and external to the set of monitored addresses.  Because the CAIDA Telescope network addresses are a darkspace, only the upper left (external $\rightarrow$ internal) quadrant will have data.

During 2020 over 20,000,000,000,000 unique packets were collected by the CAIDA Telescope.  This data set represents the largest ever assembled public corpus of Internet traffic, and is perhaps the largest public collection of streaming network events of any type.  Analysis of such a large network data set is computationally challenging \cite{lumsdaine2007challenges, kolda2009tensor, hilbert2011world}.  Using the combined resources of the Supercomputing Centers at UC San Diego, Lawrence Berkeley National Laboratory, and MIT, the spatial temporal structure of anonymized source-destination pairs from the CAIDA Telescope data has been analyzed leveraging prior work on massively parallel GraphBLAS and D4M hierarchical hypersparse matrices \cite{Kepner2009, kepner2011graph, kepner2018mathematics, reuther2018interactive, gadepally2018hyperscaling, kepner19streaming, kepner202075, kepner2021vertical} to reveal a wide range of scaling relations \cite{kepner2021spatial}.   For this study 5 contiguous subsets of $2^{30}$ CAIDA Telescope packets were selected and formed into GraphBLAS hypersparse traffic matrices at approximately 6-week intervals (see Table~\ref{tab:GreyNoise-CAIDA-Data}).  Prior work has shown that constant packet, variable time samples simplify the statistical analysis of the heavy-tail distributions commonly found in network traffic quantities \cite{kepner19hypersparse, nair2020fundamentals, kepner2022new}.  Within each of these $2^{30}$ packet windows there are 500,000 to 800,000 unique sources.  

The GreyNoise honeyfarm consists of hundreds of servers that passively collect packets from hundreds of thousands of IPs seen scanning the internet every day.  GreyNoise servers converse with these sources and  analyze and enrich these observations to identify behavior, methods and intent.  The commercial goal of GreyNoise is to analyze and label data on IPs that saturate security tools with noise. This  perspective helps analysts ignore irrelevant or harmless activity, creating more time to uncover and investigate true threats.  The GreyNoise honeyfarm outpost deduces more information about sources by responding to traffic to determine its nature and so exists in both the upper left (external $\rightarrow$ internal) quadrant and the lower right (internal  $\rightarrow$ external) quadrant of the corresponding traffic matrix (see Figure~\ref{fig:GatewayTrafficMatrix}).   For this study, GreyNoise provided data over a 15-month period which has been divided into 1-month windows (see Table~\ref{tab:GreyNoise-CAIDA-Data}).  Within each of these 1-month windows there are 1,000,000 to 14,000,000 uniques sources.

Internet data  must be handled with care, and CAIDA has pioneered standard trusted data sharing best practices that include \cite{kepner2021zero}
\begin{itemize}
\item Data is made available in curated repositories
\item Using standard anonymization methods where needed: hashing, sampling, and/or simulation
\item Registration with a repository and demonstration of legitimate research need
\item Recipients legally agree to neither repost a corpus nor deanonymize data
\item Recipients can publish analysis and data examples necessary to review research
\item Recipients agree to cite the repository and provide publications back to the repository
\item Repositories can curate enriched products developed by researchers
\end{itemize}
Within the above trusted sharing framework there are three main ways that subsets of anonymized data from multiple sources can be correlated \cite{pisharody2021realizing}
\begin{enumerate}
\item If the subset is small and the risk is low, then anonymized data can be sent back to the sources for  deanonymization.
\item If the subset is small, a third common anonymization scheme can be used and the data can be sent back to the sources for them to reanonymize in the common scheme.
\item For larger sets, an anonymization transformation table  provided by the sources  allows direct mapping from anonymized data to the common scheme.
\end{enumerate}
For this work, the first approach was used. 

\section{Network Quantities}

\begin{figure}
\center{\includegraphics[width=1.0\columnwidth]{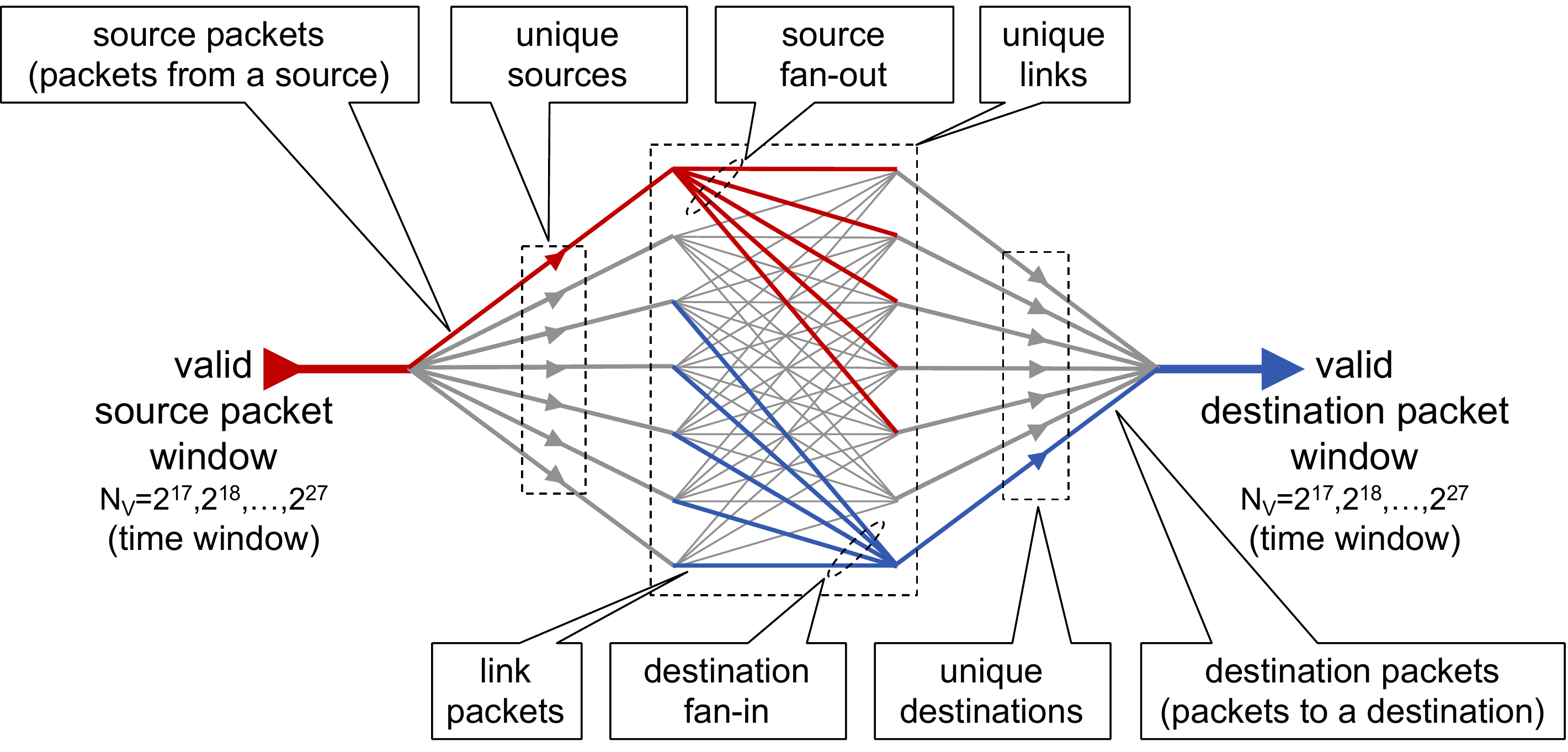}}
      	\caption{{\bf Streaming network traffic quantities.} Internet traffic streams of $N_V$ valid packets are divided into a variety of quantities for analysis: source packets, source fan-out, unique source-destination pair packets (or links), destination fan-in, and destination packets.  All of these quantities can be readily computed from anonymized hypersparse traffic matrices (see Table~\ref{tab:Aggregates}).}
      	\label{fig:NetworkDistribution}
\end{figure}

Streams of interactions between entities are found in many domains.  For Internet traffic these interactions are referred to as packets \cite{huang2018software}.  Figure~\ref{fig:NetworkDistribution} illustrates essential quantities found in all streaming dynamic networks. These quantities are all computable from anonymized traffic matrices created from the source and destinations found in packet headers.  These sources and destinations are referred as Internet Protocol (IP) addresses.

\begin{table}
\caption{Network Quantities from Traffic Matrices}
\vspace{-0.25cm}
Formulas for computing network quantities from the traffic matrix ${\bf A}_t$ at time $t$ in both summation and matrix notation. ${\bf 1}$ is a column vector of all 1's, $^{\sf T}$  is the transpose operation, and $|~|_0$ is the zero-norm that sets each nonzero value of its argument to 1\cite{karvanen2003measuring}.  These formulas are unaffected by matrix permutations and will work on anonymized data and are readily computed using  GraphBLAS hypersparse matrices or D4M associative arrays  \cite{davis18algorithm, kepner2018mathematics}.
\begin{center}
\begin{tabular}{p{1.45in}p{0.9in}p{0.6in}}
\hline
{\bf Aggregate} & {\bf ~~~~Summation} & {\bf ~Matrix} \\
{\bf Property} & {\bf ~~~~~~Notation} & {\bf Notation} \\
\hline
Valid packets $N_V$ & $~~\sum_i ~ \sum_j ~ {\bf A}_t(i,j)$ & $~{\bf 1}^{\sf T} {\bf A}_t {\bf 1}$ \\
Unique links & $~~\sum_i ~ \sum_j |{\bf A}_t(i,j)|_0$  & ${\bf 1}^{\sf T}|{\bf A}_t|_0 {\bf 1}$ \\
Link packets from $i$ to $j$ & $~~~~~~~~~~~~~~{\bf A}_t(i,j)$ & ~~~$~{\bf A}_t$ \\
Max link packets ($d_{\rm max}$) & $~~~~~\max_{ij}{\bf A}_t(i,j)$ & $\max({\bf A}_t)$ \\
\hline
Unique sources & $~\sum_i |\sum_j ~ {\bf A}_t(i,j)|_0$  & ${\bf 1}^{\sf T}|{\bf A}_t {\bf 1}|_0$ \\
Packets from source $i$ & $~~~~~~~\sum_j ~ {\bf A}_t(i,j)$ & ~~$~~{\bf A}_t  {\bf 1}$ \\
Max source packets ($d_{\rm max}$)  & $ \max_i \sum_j ~ {\bf A}_t(i,j)$ & $\max({\bf A}_t {\bf 1})$ \\
Source fan-out from $i$ & $~~~~~~~~~~\sum_j |{\bf A}_t(i,j)|_0$  & ~~~$|{\bf A}_t|_0 {\bf 1}$ \\
Max source fan-out ($d_{\rm max}$) & $ \max_i \sum_j |{\bf A}_t(i,j)|_0$  & $\max(|{\bf A}_t|_0 {\bf 1})$ \\
\hline
Unique destinations & $~\sum_j |\sum_i ~ {\bf A}_t(i,j)|_0$ & $|{\bf 1}^{\sf T} {\bf A}_t|_0 {\bf 1}$ \\
Destination packets to $j$ & $~~~~~~~\sum_i ~ {\bf A}_t(i,j)$ & ${\bf 1}^{\sf T}|{\bf A}_t|_0$ \\
Max destination packets ($d_{\rm max}$) & $ \max_j \sum_i ~ {\bf A}_t(i,j)$ & $\max({\bf 1}^{\sf T}|{\bf A}_t|_0)$ \\
Destination fan-in to $j$ & $~~~~~~~~~~\sum_i |{\bf A}_t(i,j)|_0$ & ${\bf 1}^{\sf T}~{\bf A}_t$ \\
Max destination fan-in ($d_{\rm max}$) & $ \max_j \sum_i |{\bf A}_t(i,j)|_0$ & $\max({\bf 1}^{\sf T}~{\bf A}_t)$ \\
\hline
\end{tabular}
\end{center}
\label{tab:Aggregates}
\end{table}%

The network quantities depicted in Figure~\ref{fig:NetworkDistribution} are computable from anonymized origin-destination matrices that are widely used to represent network traffic \cite{soule2004identify, zhang2005estimating, mucha2010community, tune2013internet}.  It is common to filter the packets down to a valid set for  any particular analysis.   Such filters may limit particular sources, destinations, protocols, and time windows. To reduce statistical fluctuations, the streaming data should be partitioned so that for any chosen time window all data sets have the same number of valid packets \cite{kepner19streaming}.  At a given time $t$, $N_V$ consecutive valid packets are aggregated from the traffic into a sparse matrix ${\bf A}_t$, where ${\bf A}_t(i,j)$ is the number of valid packets between the source $i$ and destination $j$. The sum of all the entries in ${\bf A}_t$ is equal to $N_V$
$$
    \sum_{i,j} {\bf A}_t(i,j) = N_V
$$
All the network quantities depicted in Figure~\ref{fig:NetworkDistribution} can be readily computed from ${\bf A}_t$ using the formulas listed in Table~\ref{tab:Aggregates}.  Because matrix operations are generally invariant to permutation (reordering of the rows and columns), these quantities can readily be computed from anonymized data using  GraphBLAS hypersparse matrices or D4M associative arrays \cite{davis18algorithm, kepner2018mathematics}.

\begin{figure}
\includegraphics[width=\columnwidth]{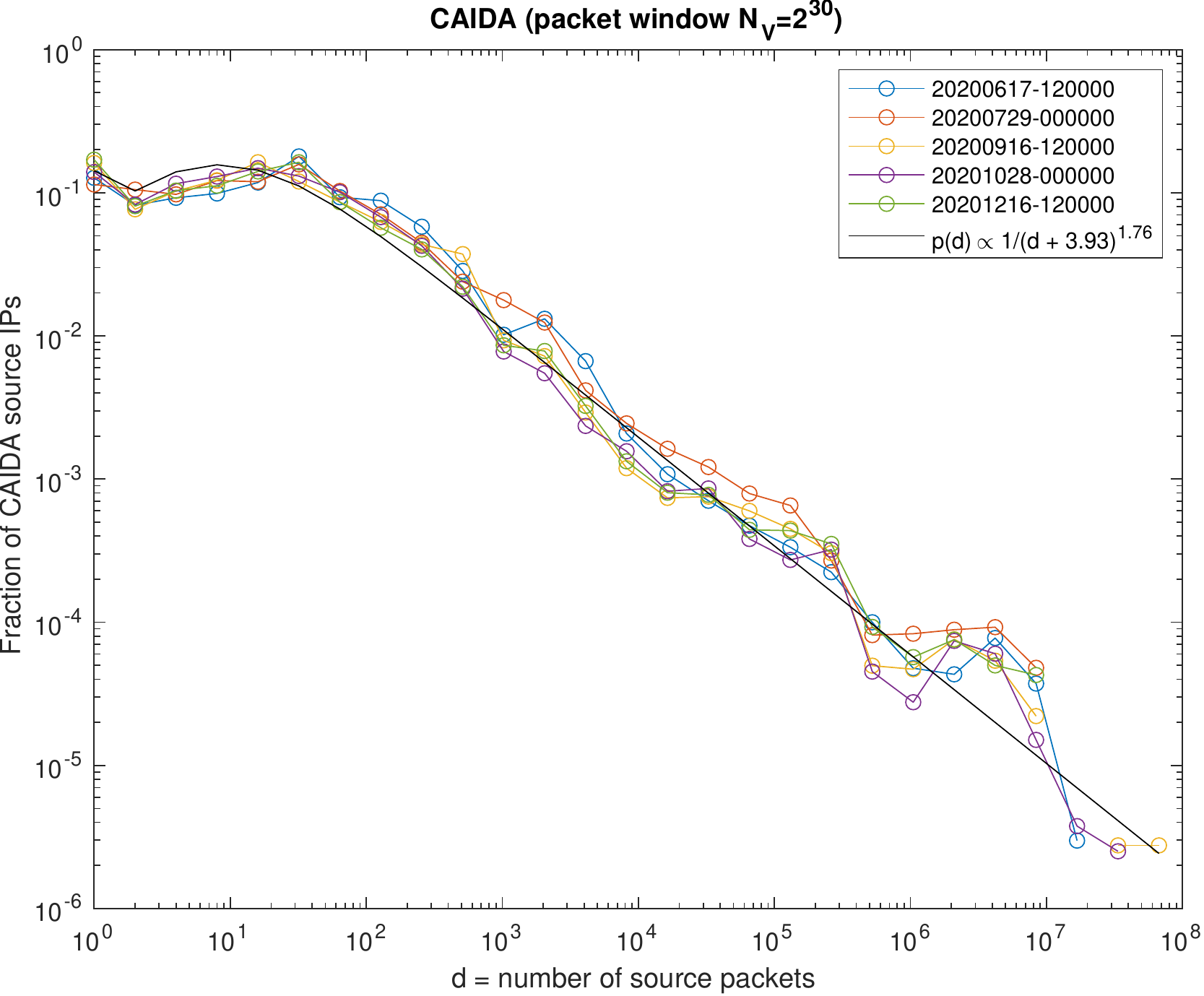}
      	\caption{{\bf CAIDA Source Packet Degree Distribution}. Differential cumulative probability (normalized histogram) for  the number (degree) of source packets from each source using logarithmic bins $d_i = 2^i$ for  $N_V = 2^{30}$ packet CAIDA samples collected over several months and at different times of day.  The observed power-law distribution can be approximated by the two parameter Zipf-Mandelbrot distribution $p(d) \propto 1/(d + \delta_{\rm zm})^{\alpha_{\rm zm}}$.}
      	\label{fig:DegreeDistribution}
\end{figure}

Processing the large volumes of data from observatories like the CAIDA Telescope requires additional computational innovations. The advent of GraphBLAS hypersparse hierarchical traffic matrices has enabled the processing of hundreds of billions of packets in minutes \cite{kepner16mathematical, buluc17design, davis18algorithm, kepner202075}. The CAIDA Telescope archives its trillions of collected packets at the supercomputing center at Lawrence Berkeley National Laboratory (LBNL) where the packets are aggregated into CryptoPAN \cite{fan2004prefix} anonymized GraphBLAS traffic matrices of $N_V = 2^{17}$ valid contiguous packets.   The $N_V = 2^{30}$ traffic matrices used in this study are constructed by hierarchically summing $2^{13}$ of these traffic smaller matrices.

Because of the large volume of CAIDA Telescope data, traffic matrices were constructed using ${2^{32}}{\times}{2^{32}}$ hypersparse GraphBLAS matrices using uint32 indices and floating point values, so 3 packets from IPv4 source 1.1.1.1 to IPv4 destination 2.2.2.2 in time-window $t$ would be represented as
$$
    {\bf A}_t(16843009,33686018) = 3.0
$$
The GreyNoise data was smaller and contains additional metadata represented as strings, so the GreyNoise data was represented using D4M associative arrays, which for the aforementioned example would be 
$$
    {\bf A}_t({\textrm{\textquotesingle1.1.1.1\textquotesingle}},{\textrm{\textquotesingle2.2.2.2\textquotesingle}}) = {\textrm{\textquotesingle3\textquotesingle}}
$$
\noindent After the unique sources and packet counts are computed from the CAIDA Telescope GraphBLAS matrices, the reduced results are converted to D4M associative arrays to facilitate correlation with the GrayNoise D4M associative arrays.

Each network quantity computed from ${\bf A}_t$ will produce a distribution of values whose magnitude is often called the degree $d$. The corresponding histogram of the network quantity is denoted by $n_t(d)$.  The largest observed value in the distribution is denoted  $d_{\rm max}$.  The normalization factor of the distribution is given by
$$
    \sum_d n_t(d)
$$
with corresponding probability
$$
    p_t(d) = n_t(d)/\sum_d n_t(d)
$$
and cumulative probability
$$
    P_t(d) = \sum_{i=1,d} p_t(d)
$$
Because of the relatively large values of $d$ observed, the measured probability at large $d$ often exhibits large fluctuations. However, the cumulative probability lacks sufficient detail to see variations around specific values of $d$, so it is typical to pool the \emph{differential cumulative probability} with logarithmic bins in $d$
$$
    D_t(d_i) = P_t(d_i) - P_t(d_{i-1})
$$
where $d_i = 2^i$ \cite{clauset2009power}.  All computed probability distributions use the same binary logarithmic binning to allow for consistent statistical comparison across data sets \cite{clauset2009power, barabasi2016network}.

\begin{figure}
\includegraphics[width=\columnwidth]{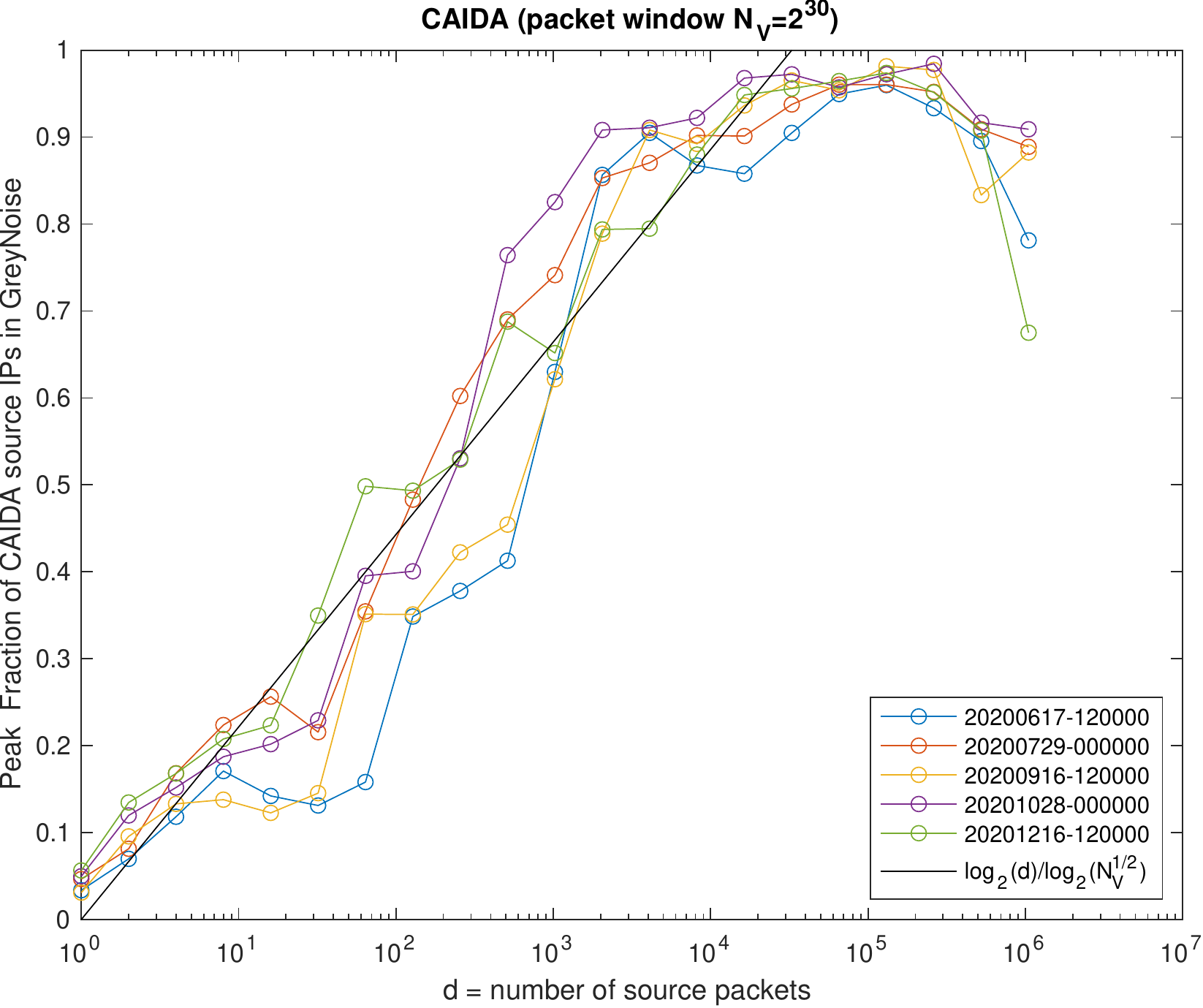}
      	\caption{{\bf Peak Correlation}.  Correlation of CAIDA source IPs with GreyNoise source IPs during the same month as a function of CAIDA source packets $d$.  CAIDA sources with $d > N_V^{1/2}$ packets in the packet window are very likely to appear in the GreyNoise data of the same month. CAIDA sources with $d < N_V^{1/2}$ appear with a probability  $\log_2(d)/\log_2(N_V^{1/2})$.}
      	\label{fig:GreynoiseCAIDA-peak-total-dN}
\end{figure}

Figure~\ref{fig:DegreeDistribution} shows the distribution of external $\rightarrow$ internal source packets for 5 CAIDA samples with $N_V = 2^{30}$.  The resulting distribution has the power-law shape frequently observed in network  data  \cite{leland1994self, faloutsos1999power, albert1999internet, barabasi1999emergence, adamic2000power, barabasi2009scale, mahanti2013tale}.   The power-law distribution in Figure~\ref{fig:DegreeDistribution} can be approximated by the two parameter Zipf-Mandelbrot distribution that is widely seen in network data \cite{kepner19hypersparse, kepner2022new}
$$
   p(d) \propto 1/(d + \delta_{\rm zm})^{\alpha_{\rm zm}}
$$

\section{Correlation Results}

  An important objective of correlating observations from different locations is to determine how the observations are similar and different.  A first step is to ask what fraction of the CAIDA Telescope sources are also seen in the GreyNoise observations during the same month.  Figure~\ref{fig:GreynoiseCAIDA-peak-total-dN}  plots the fraction of CAIDA sources seen in the GrayNoise data as a function of the number of CAIDA source packets $d$ binned logarithmically.   Figure~\ref{fig:GreynoiseCAIDA-peak-total-dN}  shows that bright CAIDA sources with $d > N_V^{1/2} = 2^{15}$ are nearly always also seen by the GreyNoise observations during the same month.  Likewise, for fainter sources with $d < N_V^{1/2}$ packets the probability of the CAIDA source being seen in the GreyNoise data can be empirically  approximated as
$$
   p(d ~ | ~ {\rm CAIDA  ~ \& ~ GreyNoise}) \approx \log_2(d)/\log_2(N_V^{1/2})
$$

\begin{figure}
\includegraphics[width=\columnwidth]{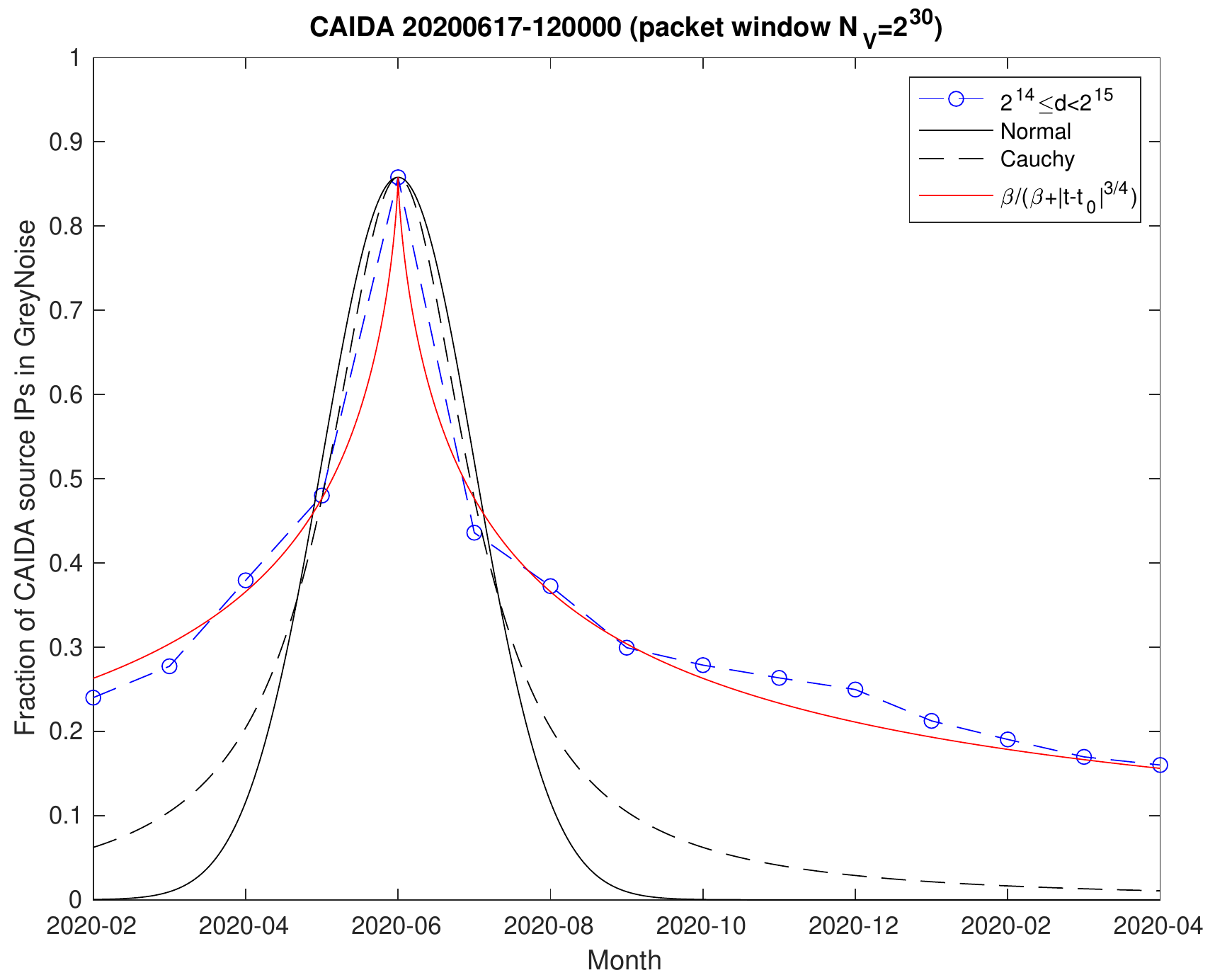}
      	\caption{{\bf Temporal Correlation}. Fraction of CAIDA 2020-06-17 sources with $2^{14} \leq$ source packets $<2^{15}$ found in GreyNoise sources over a 15-month period.  Data are fit to Gaussian, Cauchy, and the modified Cauchy distribution $\beta/(\beta + |t-t_0|^\alpha)$.}
      	\label{fig:CAIDA-20200617-GreyNoise-d14}
\end{figure}

Another important comparison is how the correlations change as the time between CAIDA and GreyNoise measurements increases.  Figure~\ref{fig:CAIDA-20200617-GreyNoise-d14} shows the correlation of CAIDA sources with $2^{14} \leq d < 2^{15}$ packets with GreyNoise data over a 15 month span.  The correlation between the CAIDA and GreyNoise sources drops quickly and then levels off to a background level.  The data have been fit to Gaussian (Normal), Cauchy \cite{stigler1974studies, larson1981urban}, and modified Cauchy distributions.  Specifically, the following function that we will refer to as the modified Cauchy distribution
$$
   f_{\rm Cauchy}^{\rm modified}(t) \propto \frac{\beta}{\beta + |t - t_0|^\alpha}
$$
where $t_0$ is the CAIDA measurement time, $t$ is the GreyNoise measurement time, with exponent $\alpha > 0$, and scale factor $\beta > 0$.  Setting $\alpha = 2$ and $\beta = \gamma^\alpha$ results in the standard Cauchy distribution
$$
   f_{\rm Cauchy}(t) \propto \frac{\gamma^2}{\gamma^2 + |t - t_0|^2}
$$
Figure~\ref{fig:CAIDA-20200617-GreyNoise-d14} is well approximated by the modified Cauchy distribution.

\begin{figure*}
\includegraphics[width=0.7\columnwidth]{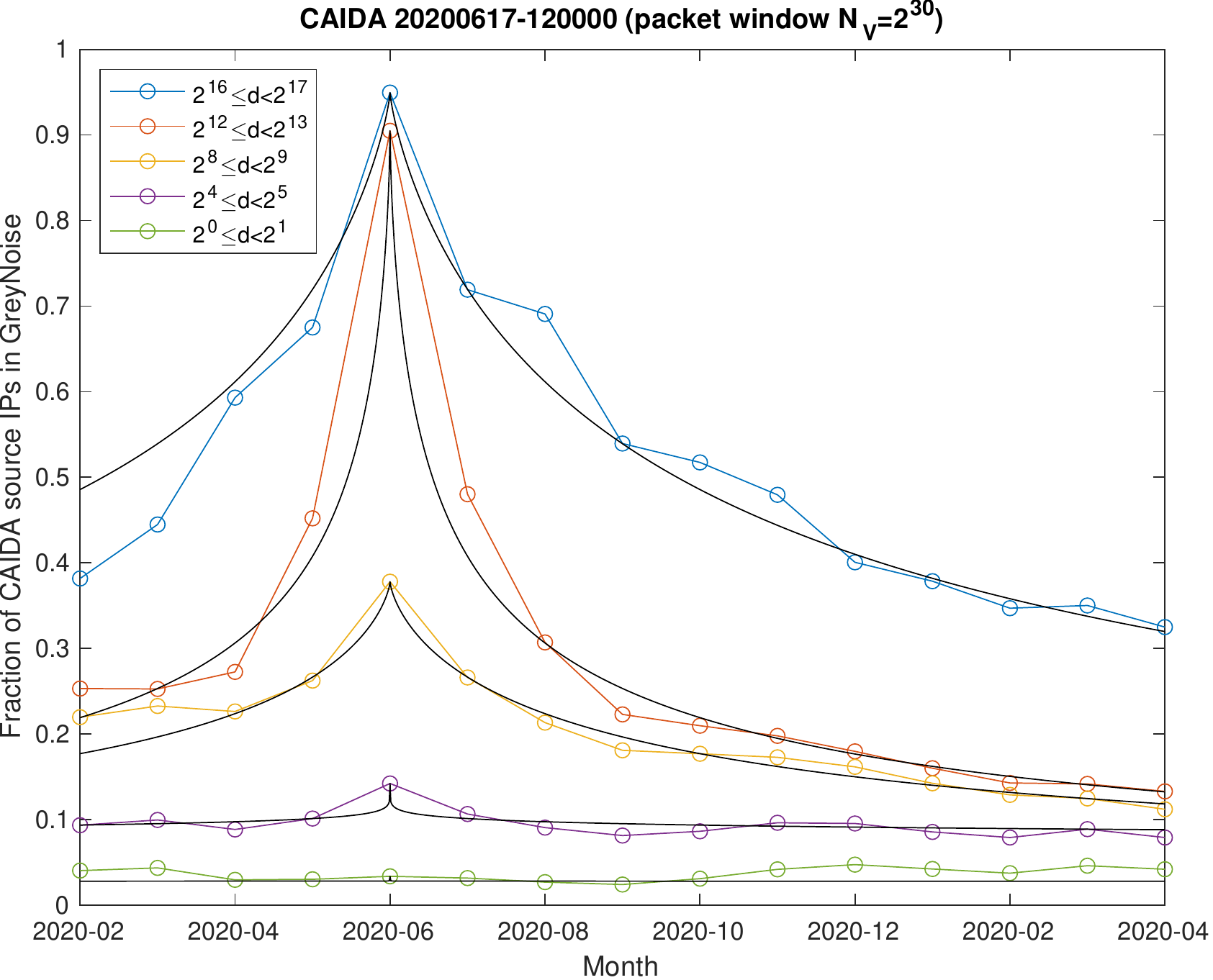}
\includegraphics[width=0.7\columnwidth]{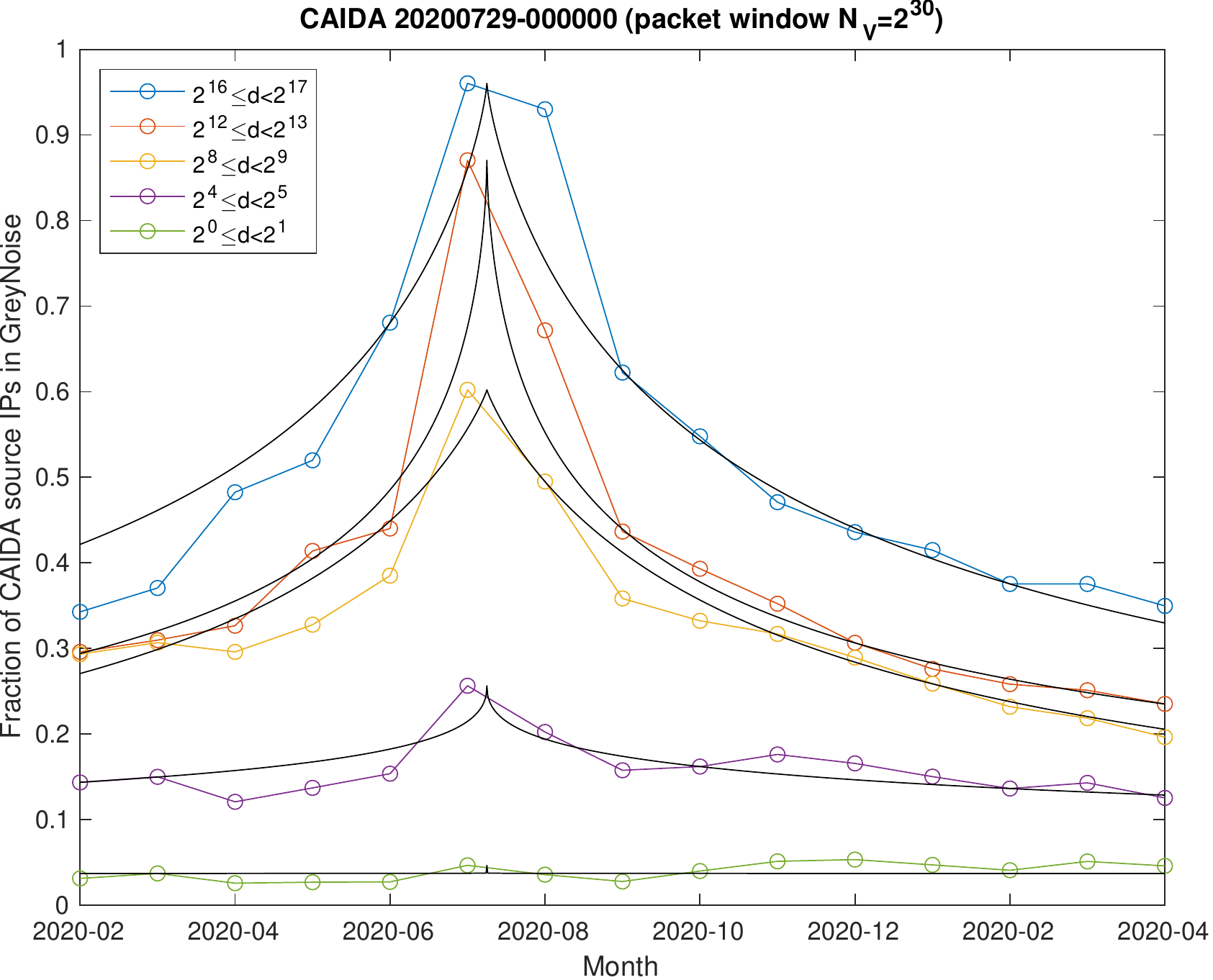}
\includegraphics[width=0.7\columnwidth]{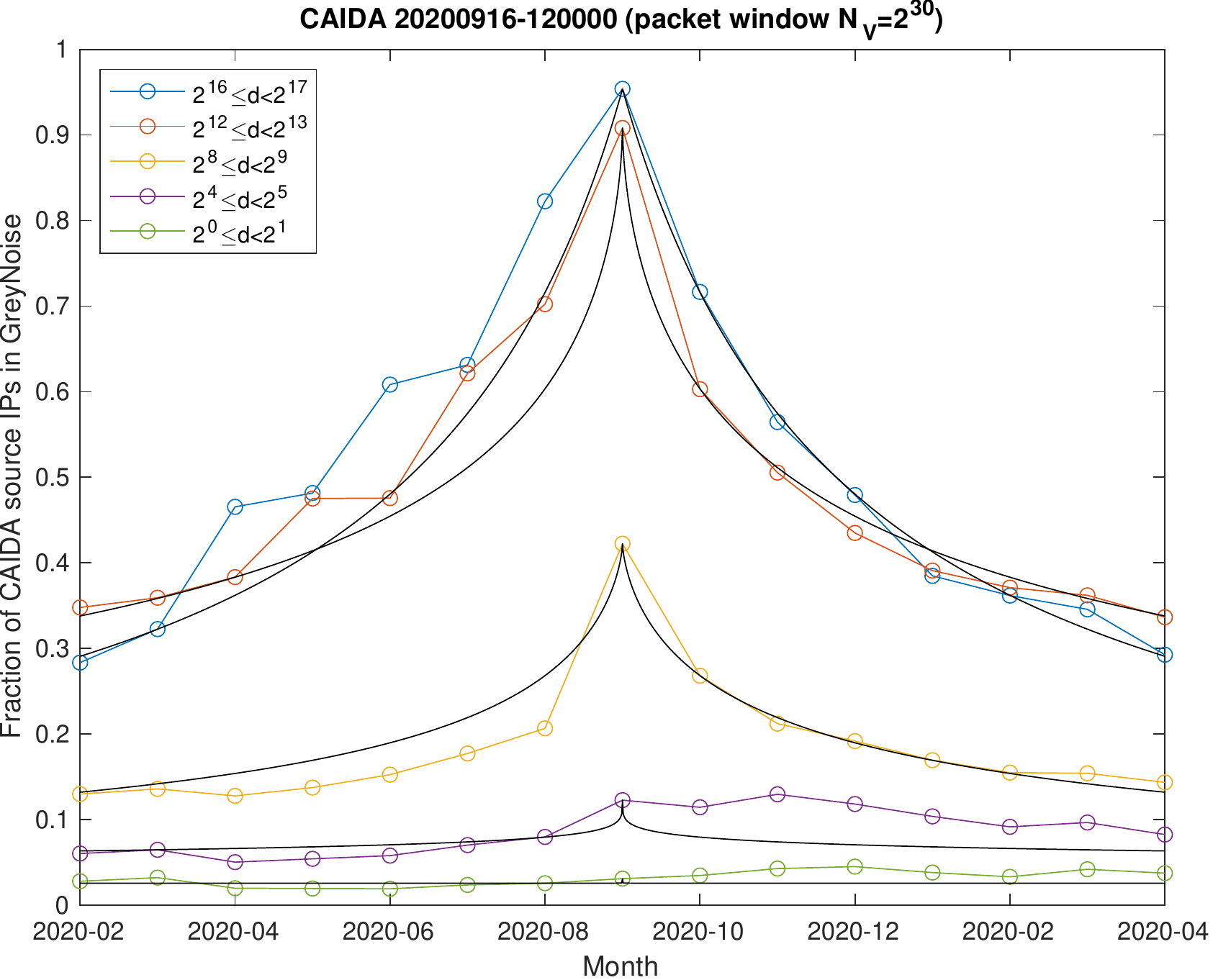}
\includegraphics[width=0.7\columnwidth]{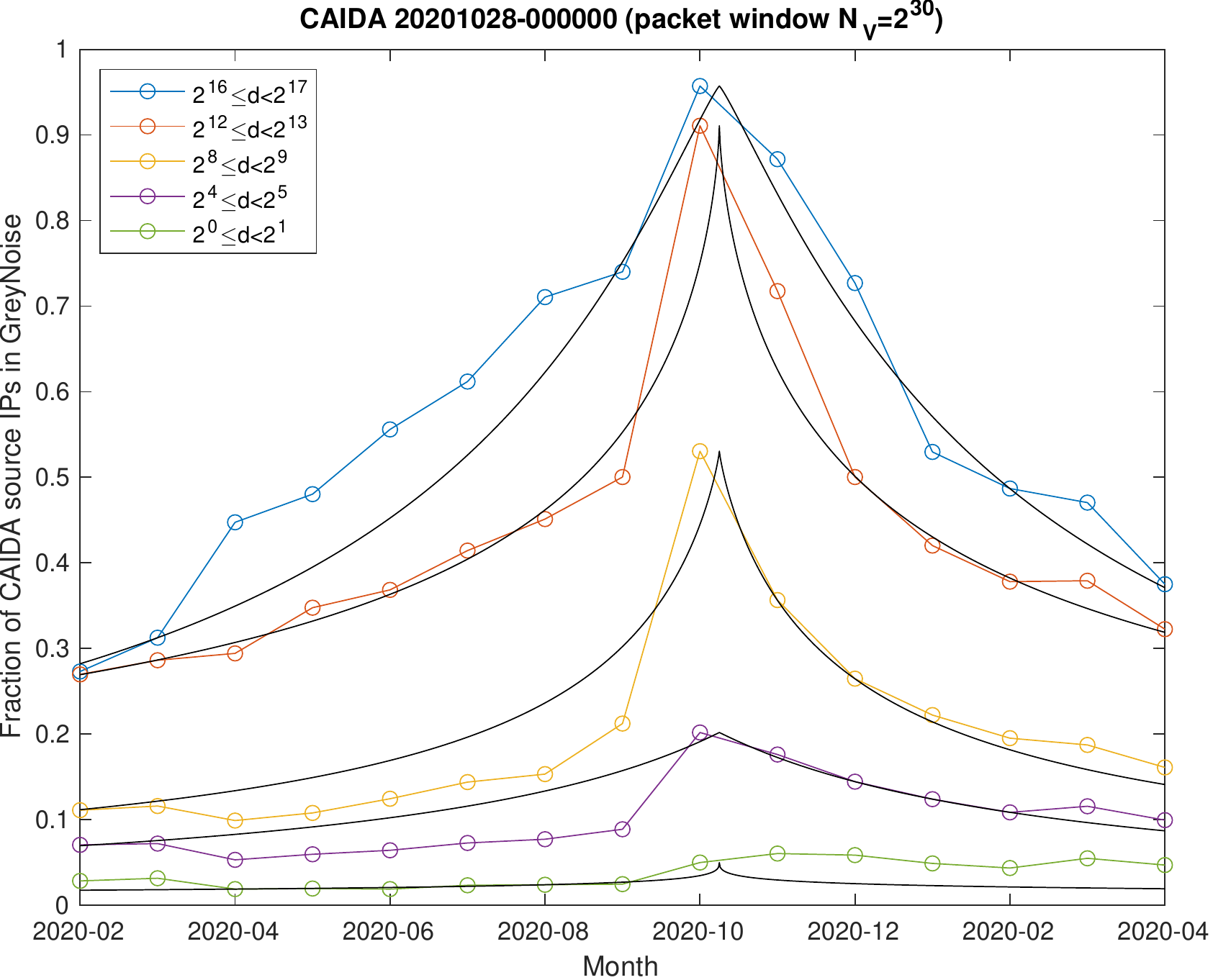}
\includegraphics[width=0.7\columnwidth]{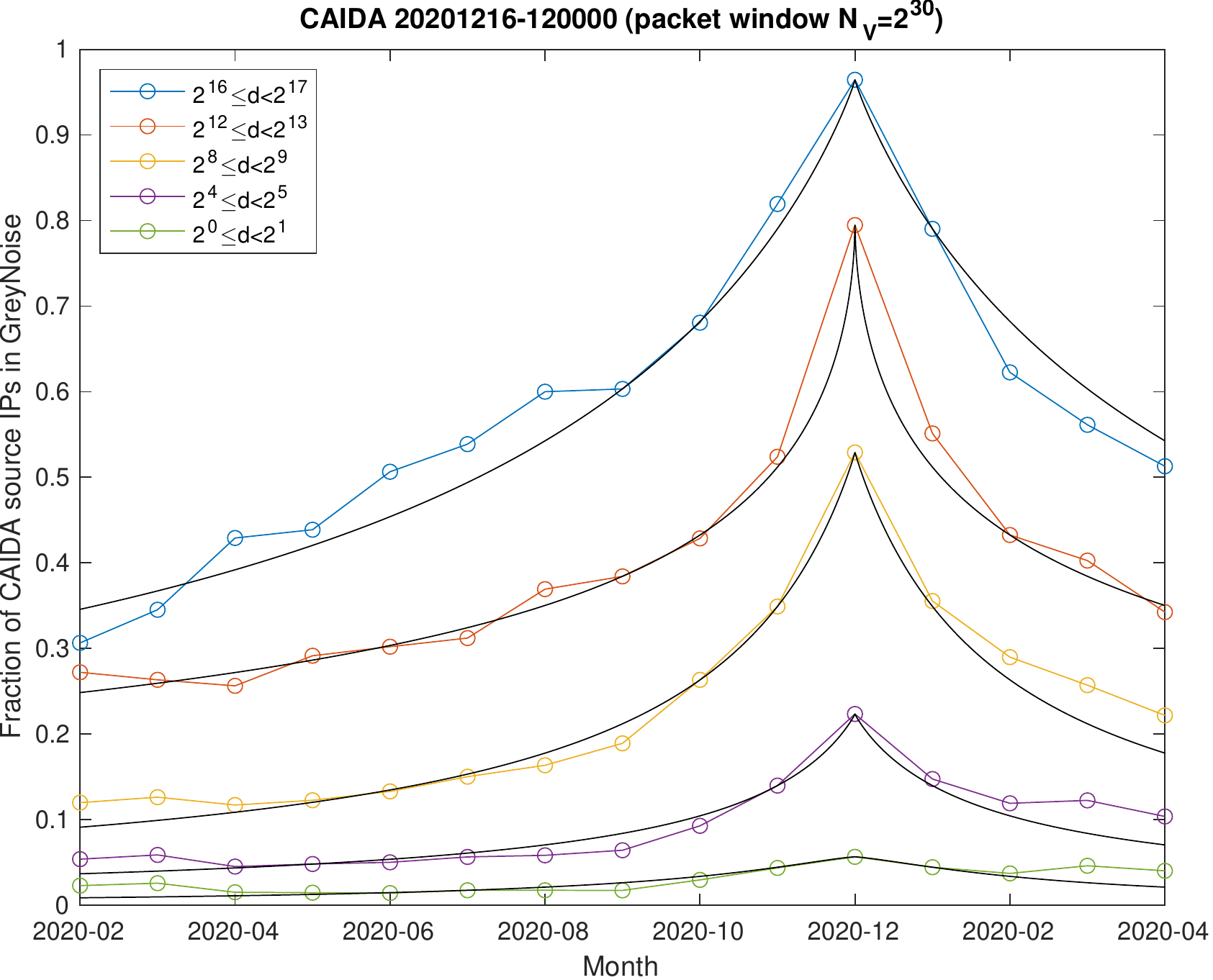}
   \caption{{\bf Temporal Correlation and Packet Degree}.  Fraction of CAIDA sources found in GreyNoise sources over a 15-month period for $d \leq$ source packets $<2d$.  Black lines show the best fit modified Cauchy distributions $\beta/(\beta + |t-t_0|^\alpha$).}
\label{fig:GreyNoiseCAIDA}
\end{figure*}

\begin{figure}
\includegraphics[width=1.0\columnwidth]{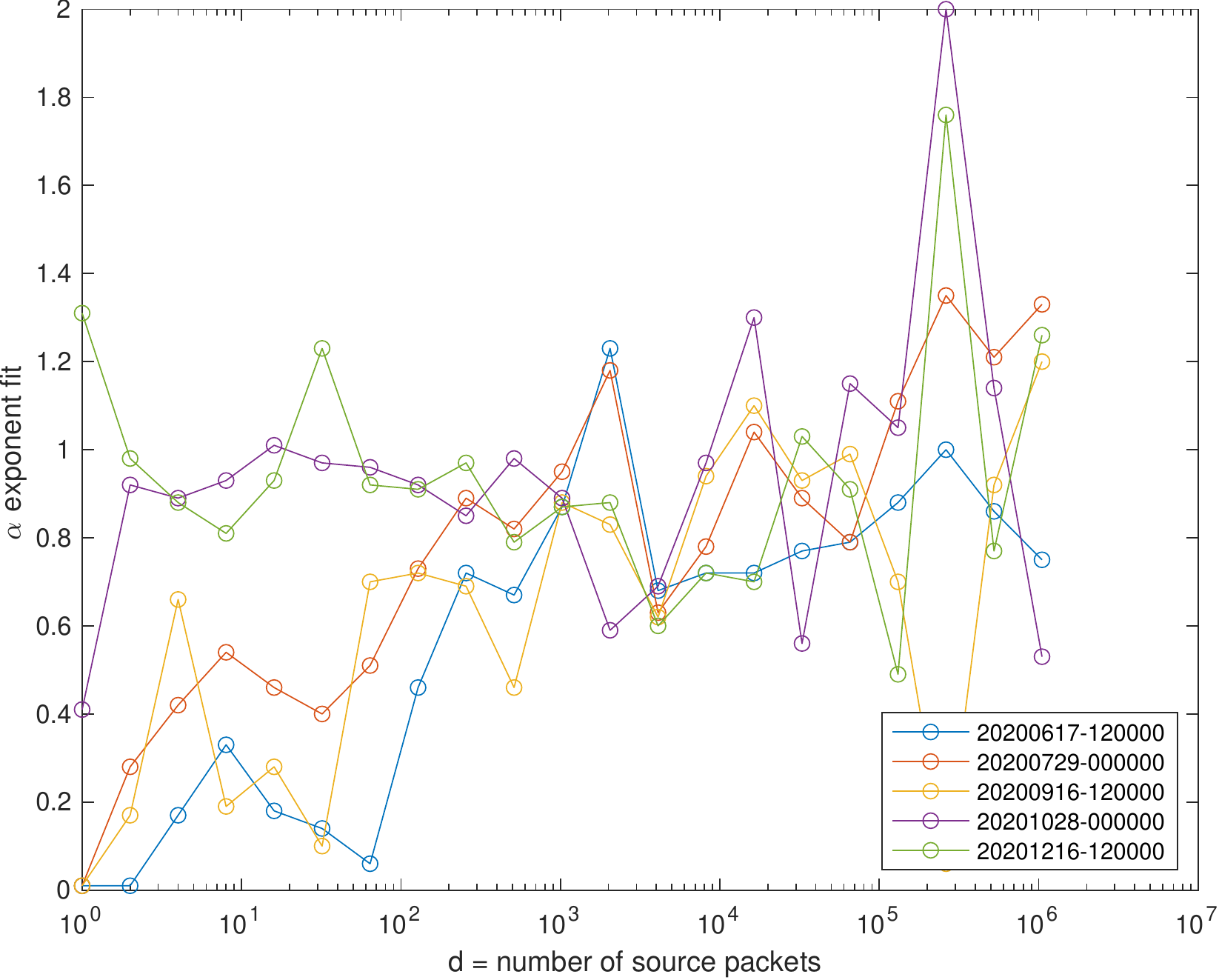}
   \caption{{\bf Modified Cauchy Distribution $\alpha$}.  Best fit $\alpha$ parameter from the modified Cauchy distribution models as a function of the number of CAIDA source packets $d$.
 }
   \label{fig:AlphaBeta}
\end{figure}

\begin{figure}
\includegraphics[width=1.0\columnwidth]{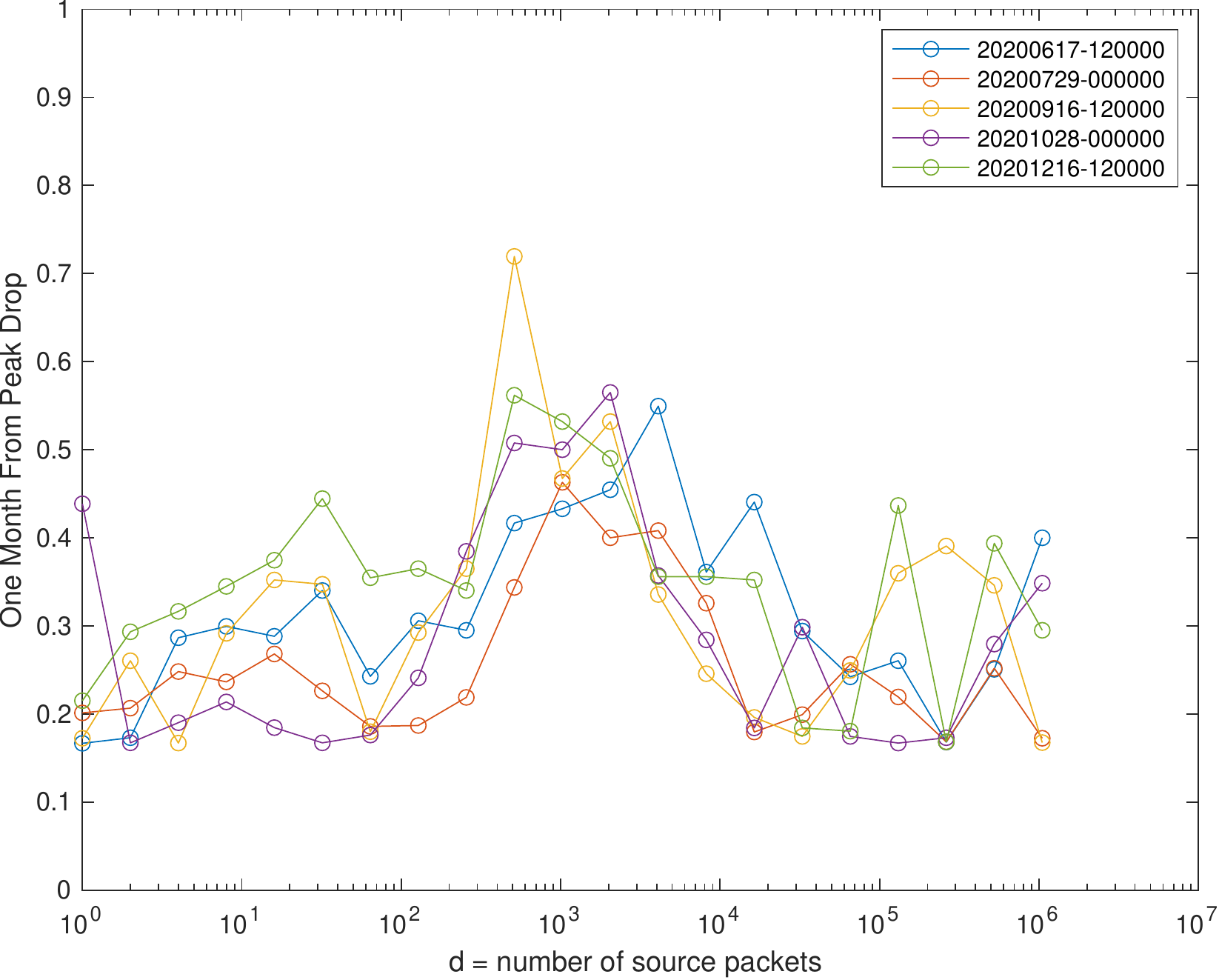}
   \caption{{\bf One Month Drop}.  One month from peak drop $1/(\beta + 1)$ as derived from the $\beta$ fit parameter of the modified Cauchy distributions as a function of the number of CAIDA source packets $d$.
 }
   \label{fig:OMFPD}
\end{figure}

Figure~\ref{fig:GreyNoiseCAIDA} shows the CAIDA GreyNoise temporal correlations for all the CAIDA samples for selected source packet ranges.  All the curves are fit to the modified Cauchy distribution by generating all distributions over a range of possible $\alpha$ and $\beta$ values, normalizing to the peak in the data, and then selecting the $\alpha$ and $\beta$ that minimize the $|~|^{1/2}$ norm.  The best-fit scaling exponent $\alpha$ for all source packet ranges is shown in Figure~\ref{fig:AlphaBeta}.  The quantity $1 - \beta/(\beta + 1) = 1/(\beta + 1)$ provides the relative one month drop from the peak and is shown in Figure~\ref{fig:OMFPD}.

\section{Discussion}

Understanding the baseline statistical distributions of traffic  are essential to the scientific understanding of the Internet.  These data lend themselves to a number of observations about the statistical distributions of Internet traffic, the stability of these distributions over time, the correlation of measurements from different locations, and mathematical models of these approximations. Each of these observations provides a basis for predictions for future measurements and for theoretical modeling of the underlying generative processes.

Figure~\ref{fig:DegreeDistribution} shows that the source packet distributions of the CAIDA samples collected at different times have similar statistical distributions  with small variations.  Furthermore, the packet distribution is well approximated by a Zipf-Mandelbrot distribution.  The temporal consistency of the observations with a stable Zipf-Mandelbrot model agrees with prior observations of the CAIDA Telescope \cite{kepner2021spatial}, the CAIDA Chicago A \& B Internet traffic collection \cite{kepner19hypersparse, kepner2022new}, the MAWI Internet traffic collection \cite{kepner19hypersparse, kepner2022new}, and other network gateways \cite{kepner2020multi}.  These observations have led to the development of new generative models of network traffic that extend prior preferential attachment models with parameters to describe adversarial traffic \cite{devlin2021hybrid}.

Figure~\ref{fig:GreynoiseCAIDA-peak-total-dN}  shows that the temporal statistical consistency can extend to the correlation of sources seen at separate locations. It also suggests that sources above a certain brightness are very likely to be seen, in contrast to prior observations \cite{nawrocki2021far}.  In this case sources brighter than $d > N_V^{1/2}$ packets or whose fraction of the total packets is greater than $N_V^{-1/2}$.  Below this threshold the probability of being seen in both CAIDA and GreyNoise during the same month is approximated by $\log_2(d)/\log_2(N_V^{1/2})$.  This purely empirical logarithmic distribution and the role of the empirical value $N_V^{1/2}$ should be tested with additional comparative observations.  Perhaps $N_V^{1/2}$ is connected to the fact that the number of unique sources seen at the CAIDA Telescope and other locations  is approximately proportional to  $N_V^{1/2}$ \cite{kepner2020multi, kepner2021spatial}.   Likewise the logarithmic distribution could be an interesting target for new theoretical models. 

Figures~\ref{fig:CAIDA-20200617-GreyNoise-d14} and \ref{fig:GreyNoiseCAIDA} show the temporal correlation between the GreyNoise and CAIDA sources of various brightness showing that the temporal statistical consistency extends  over significant time.  The correlation as a function of time is well approximated by the modified Cauchy distribution.  While it is certainly expected that brighter sources seen at one location on the Internet are more likely to be seen in another location at the same time, the simple empirical relation connecting the CAIDA and GreyNoise observations is intriguing.  A common geometric interpretation of the Cauchy distribution is the probability of a randomly blinking rotating beam positioned above the point $t_0$ at a distance $\gamma$ from a wall hitting a point $t$ on a wall.  Such a geometric analogy may represent possible direction for theoretical exploration.

Figure~\ref{fig:AlphaBeta} shows the best fit $\alpha$-exponent from the modified Cauchy distribution as a function of CAIDA source packets.  These observations suggest that 1 is a typical value of $\alpha$.  The $\beta$ scale factor is shown in Figure~\ref{fig:OMFPD} via the expression $1/(\beta + 1)$ which measures the relative 1-month drop off of the modified Cauchy distributions.  The typical 1-month drop off is above 20\% and increases to 50\% for $d \approx 10^3$ source packets.    These observations suggest the modified Cauchy distributions for source packets around $d \approx10^3$ are typically
$$
    f_{\rm Cauchy}^{\rm modified}(t; d \approx 10^3) \propto \frac{1}{1 + |t - t_0|}
$$
For other values of source packet a typical modified Cauchy distribution would be
$$
    f_{\rm Cauchy}^{\rm modified}(t; d \not\approx 10^3 ) \propto \frac{4}{4 + |t - t_0|}
$$
These empirical observations offer potential starting points for further theoretical observations.

\section{Conclusions and Future Work}

Scientific exploration of the Internet now requires endeavors akin to those used to understand the land, sea, air, and space environments. Understanding  what is the normal statistical behavior of Internet traffic  is a critical first step. Comparing observations  from different locations on the Internet is an effective means for determining which network quantities vary or change.   Using data from the largest Internet telescope (the CAIDA darknet telescope) and a commercial outpost (the GreyNoise honeyfarm) this  work explores the correlation of the sources seen using GraphBLAS hyperspace matrices and D4M associative arrays.  The  CAIDA sources are well approximated by a Zipf-Mandelbrot distribution.  Over a 6-month period 70\% of the brightest (highest frequency) sources in the CAIDA telescope are consistently detected by coeval observations in the GreyNoise honeyfarm.  This overlap drops as the sources dim (reduce frequency) and as the time difference between the observations grows.  The probability of seeing a CAIDA source is proportional to the logarithm of the brightness.  The temporal correlations are well described by a modified Cauchy distribution.  These observations are consistent with a correlated high frequency beam of sources that drifts on time scales of months.  Each of these observations provides a basis for predictions for future measurements and for theoretical modeling of the underlying generative processes.

\section*{Acknowledgments}

The authors wish to acknowledge the following individuals for their contributions and support: Bob Bond, Ronisha Carter, Cary Conrad, Alan Edelman, Tucker Hamilton, Jeff Gottschalk, Nathan Frey, Chris Hill, Mike Kanaan, Tim Kraska, Andrew Morris, Charles Leiserson, Dave Martinez, Mimi McClure, Joseph McDonald, Christian Prothmann, John Radovan, Steve Rejto, Daniela Rus, Allan Vanterpool, Adam Weirman, Matthew Weiss, Marc Zissman.

\bibliographystyle{ieeetr}
\bibliography{CrossCorrelation}

\begin{thebibliography}{10}

\bibitem{Cisco2018-2023}
``{\it Cisco Visual Networking Index: Forecast and Trends, 2018–2023}.''
  https://www.cisco.com/c/en/us/solutions/collateral/executive-perspectives/annual-internet-report/white-paper-c11-741490.html.

\bibitem{kepner2021zero}
J.~Kepner, J.~Bernays, S.~Buckley, K.~Cho, C.~Conrad, L.~Daigle, K.~Erhardt,
  V.~Gadepally, B.~Greene, M.~Jones, R.~Knake, B.~Maggs, P.~Michaleas,
  C.~Meiners, A.~Morris, A.~Pentland, S.~Pisharody, S.~Powazek, A.~Prout,
  P.~Reiner, K.~Suzuki, K.~Takhashi, T.~Tauber, L.~Walker, and D.~Stetson,
  ``Zero botnets: An observe-pursue-counter approach.'' Belfer Center Reports,
  6 2021.

\bibitem{claffy2000measuring}
K.~Claffy, ``Measuring the internet,'' {\em IEEE Internet Computing}, vol.~4,
  no.~1, pp.~73--75, 2000.

\bibitem{li2013survey}
B.~Li, J.~Springer, G.~Bebis, and M.~H. Gunes, ``A survey of network flow
  applications,'' {\em Journal of Network and Computer Applications}, vol.~36,
  no.~2, pp.~567--581, 2013.

\bibitem{rabinovich2016measuring}
M.~Rabinovich and M.~Allman, ``Measuring the internet,'' {\em IEEE Internet
  Computing}, vol.~20, no.~4, pp.~6--8, 2016.

\bibitem{ClaffyClark2020}
k.~claffy and D.~Clark, ``Workshop on internet economics (wie 2019) report,''
  {\em SIGCOMM Comput. Commun. Rev.}, vol.~50, p.~53–59, May 2020.

\bibitem{CAIDA2019}
``{\it CAIDA Anonymized Internet Traces Dataset (April 2008 - January 2019)}.''
  https://www.caida.org/catalog/datasets/passive\_dataset/.

\bibitem{CAIDA2022}
``{\it UCSD Network Telescope}.''
  https://www.caida.org/projects/network\_telescope/.

\bibitem{GCA2022}
``{\it Global Cyber Alliance}.'' https://www.globalcyberalliance.org/.

\bibitem{Greynoise2022}
``{\it Greynoise}.'' https://greynoise.io/.

\bibitem{MAWI2020}
``{\it MAWI Working Group Traffic Archive}.'' http://mawi.wide.ad.jp/mawi/.

\bibitem{Shadowserver2022}
``{\it Shadowserver Foundation}.'' https://www.shadowserver.org/.

\bibitem{kepner2020multi}
J.~Kepner, C.~Meiners, C.~Byun, S.~McGuire, T.~Davis, W.~Arcand, J.~Bernays,
  D.~Bestor, W.~Bergeron, V.~Gadepally, R.~Harnasch, M.~Hubbell, M.~Houle,
  M.~Jones, A.~Kirby, A.~Klein, L.~Milechin, J.~Mullen, A.~Prout, A.~Reuther,
  A.~Rosa, S.~Samsi, D.~Stetson, A.~Tse, C.~Yee, and P.~Michaleas,
  ``Multi-temporal analysis and scaling relations of 100,000,000,000 network
  packets,'' in {\em 2020 IEEE High Performance Extreme Computing Conference
  (HPEC)}, pp.~1--6, 2020.

\bibitem{marder2021inferring}
A.~Marder, K.~C. Claffy, and A.~C. Snoeren, ``Inferring cloud interconnections:
  Validation, geolocation, and routing behavior,'' in {\em International
  Conference on Passive and Active Network Measurement}, pp.~230--246,
  Springer, 2021.

\bibitem{luckie2021learning}
M.~Luckie, B.~Huffaker, A.~Marder, Z.~Bischof, M.~Fletcher, and K.~Claffy,
  ``Learning to extract geographic information from internet router
  hostnames,'' in {\em Proceedings of the 17th International Conference on
  emerging Networking EXperiments and Technologies}, pp.~440--453, 2021.

\bibitem{carisimo2021identifying}
E.~Carisimo, A.~Gamero-Garrido, A.~C. Snoeren, and A.~Dainotti, ``Identifying
  ases of state-owned internet operators,'' in {\em Proceedings of the 21st ACM
  Internet Measurement Conference}, pp.~687--702, 2021.

\bibitem{manousis2021shape}
A.~Manousis, H.~Shah, H.~Milner, Y.~Li, H.~Zhang, and V.~Sekar, ``The shape of
  view: an alert system for video viewership anomalies,'' in {\em Proceedings
  of the 21st ACM Internet Measurement Conference}, pp.~245--260, 2021.

\bibitem{liu2021s}
E.~Liu, G.~Akiwate, M.~Jonker, A.~Mirian, S.~Savage, and G.~M. Voelker, ``Who's
  got your mail? characterizing mail service provider usage,'' in {\em
  Proceedings of the 21st ACM Internet Measurement Conference}, pp.~122--136,
  2021.

\bibitem{yan2021detecting}
Z.~Yan, Z.~Li, W.~Bai, N.~Yu, H.~Zhu, and L.~Sun, ``Detecting internet-scale
  surveillance devices using rtsp recessive features,'' in {\em International
  Conference on Science of Cyber Security}, pp.~21--35, Springer, 2021.

\bibitem{cheng2021botnet}
H.~Cheng, Y.~Shen, T.~Cheng, Y.~Fang, and J.~Ling, ``Botnet detection based on
  multilateral attribute graph,'' in {\em International Conference on Science
  of Cyber Security}, pp.~66--76, Springer, 2021.

\bibitem{nawrocki2021far}
M.~Nawrocki, M.~Jonker, T.~C. Schmidt, and M.~W{\"a}hlisch, ``The far side of
  dns amplification: tracing the ddos attack ecosystem from the internet
  core,'' in {\em Proceedings of the 21st ACM Internet Measurement Conference},
  pp.~419--434, 2021.

\bibitem{kepner19hypersparse}
J.~{Kepner}, K.~{Cho}, K.~{Claffy}, V.~{Gadepally}, P.~{Michaleas}, and
  L.~{Milechin}, ``Hypersparse neural network analysis of large-scale internet
  traffic,'' in {\em 2019 IEEE High Performance Extreme Computing Conference
  (HPEC)}, pp.~1--11, 2019.

\bibitem{nair2020fundamentals}
J.~Nair, A.~Wierman, and B.~Zwart, ``The fundamentals of heavy tails:
  Properties, emergence, and estimation,'' {\em Preprint, California Institute
  of Technology}, 2020.

\bibitem{kepner2022new}
J.~Kepner, K.~Cho, K.~Claffy, V.~Gadepally, S.~McGuire, L.~Milechin, W.~Arcand,
  D.~Bestor, W.~Bergeron, C.~Byun, M.~Hubbell, M.~Houle, M.~Jones, A.~Prout,
  A.~Reuther, A.~Rosa, S.~Samsi, C.~Yee, and P.~Michaleas, ``New phenomena in
  large-scale internet traffic,'' in {\em Massive Graph Analytics} (D.~Bader,
  ed.), pp.~1--53, Chapman and Hall/CRC, 2022.

\bibitem{lumsdaine2007challenges}
A.~Lumsdaine, D.~Gregor, B.~Hendrickson, and J.~Berry, ``Challenges in parallel
  graph processing,'' {\em Parallel Processing Letters}, vol.~17, no.~01,
  pp.~5--20, 2007.

\bibitem{kolda2009tensor}
T.~G. Kolda and B.~W. Bader, ``Tensor decompositions and applications,'' {\em
  SIAM review}, vol.~51, no.~3, pp.~455--500, 2009.

\bibitem{hilbert2011world}
M.~Hilbert and P.~L{\'o}pez, ``The world's technological capacity to store,
  communicate, and compute information,'' {\em Science}, p.~1200970, 2011.

\bibitem{Kepner2009}
J.~Kepner, {\em Parallel MATLAB for Multicore and Multinode Computers}.
\newblock SIAM, 2009.

\bibitem{kepner2011graph}
J.~Kepner and J.~Gilbert, {\em Graph algorithms in the language of linear
  algebra}.
\newblock SIAM, 2011.

\bibitem{kepner2018mathematics}
J.~Kepner and H.~Jananthan, {\em Mathematics of big data: Spreadsheets,
  databases, matrices, and graphs}.
\newblock MIT Press, 2018.

\bibitem{reuther2018interactive}
A.~{Reuther}, J.~{Kepner}, C.~{Byun}, S.~{Samsi}, W.~{Arcand}, D.~{Bestor},
  B.~{Bergeron}, V.~{Gadepally}, M.~{Houle}, M.~{Hubbell}, M.~{Jones},
  A.~{Klein}, L.~{Milechin}, J.~{Mullen}, A.~{Prout}, A.~{Rosa}, C.~{Yee}, and
  P.~{Michaleas}, ``Interactive supercomputing on 40,000 cores for machine
  learning and data analysis,'' in {\em 2018 IEEE High Performance extreme
  Computing Conference (HPEC)}, pp.~1--6, 2018.

\bibitem{gadepally2018hyperscaling}
V.~{Gadepally}, J.~{Kepner}, L.~{Milechin}, W.~{Arcand}, D.~{Bestor},
  B.~{Bergeron}, C.~{Byun}, M.~{Hubbell}, M.~{Houle}, M.~{Jones},
  P.~{Michaleas}, J.~{Mullen}, A.~{Prout}, A.~{Rosa}, C.~{Yee}, S.~{Samsi}, and
  A.~{Reuther}, ``Hyperscaling internet graph analysis with d4m on the mit
  supercloud,'' in {\em 2018 IEEE High Performance extreme Computing Conference
  (HPEC)}, pp.~1--6, Sep. 2018.

\bibitem{kepner19streaming}
J.~{Kepner}, V.~{Gadepally}, L.~{Milechin}, S.~{Samsi}, W.~{Arcand},
  D.~{Bestor}, W.~{Bergeron}, C.~{Byun}, M.~{Hubbell}, M.~{Houle}, M.~{Jones},
  A.~{Klein}, P.~{Michaleas}, J.~{Mullen}, A.~{Prout}, A.~{Rosa}, C.~{Yee}, and
  A.~{Reuther}, ``Streaming 1.9 billion hypersparse network updates per second
  with d4m,'' in {\em 2019 IEEE High Performance Extreme Computing Conference
  (HPEC)}, pp.~1--6, 2019.

\bibitem{kepner202075}
J.~Kepner, T.~Davis, C.~Byun, W.~Arcand, D.~Bestor, W.~Bergeron, V.~Gadepally,
  M.~Hubbell, M.~Houle, M.~Jones, A.~Klein, P.~Michaleas, L.~Milechin,
  J.~Mullen, A.~Prout, A.~Rosa, S.~Samsi, C.~Yee, and A.~Reuther,
  ``75,000,000,000 streaming inserts/second using hierarchical hypersparse
  graphblas matrices,'' in {\em 2020 IEEE International Parallel and
  Distributed Processing Symposium Workshops (IPDPSW)}, pp.~207--210, 2020.

\bibitem{kepner2021vertical}
J.~Kepner, T.~Davis, C.~Byun, W.~Arcand, D.~Bestor, W.~Bergeron, V.~Gadepally,
  M.~Houle, M.~Hubbell, M.~Jones, A.~Klein, L.~Milechin, J.~Mullen, A.~Prout,
  A.~Reuther, A.~Rosa, S.~Samsi, C.~Yee, and P.~Michaleas, ``Vertical,
  temporal, and horizontal scaling of hierarchical hypersparse graphblas
  matrices,'' in {\em 2021 IEEE High Performance Extreme Computing Conference
  (HPEC)}, pp.~1--6, IEEE, 2021.

\bibitem{kepner2021spatial}
J.~Kepner, M.~Jones, D.~Andersen, A.~Buluç, C.~Byun, K.~Claffy, T.~Davis,
  W.~Arcand, J.~Bernays, D.~Bestor, W.~Bergeron, V.~Gadepally, M.~Houle,
  M.~Hubbell, A.~Klein, C.~Meiners, L.~Milechin, J.~Mullen, S.~Pisharody,
  A.~Prout, A.~Reuther, A.~Rosa, S.~Samsi, D.~Stetson, A.~Tse, C.~Yee, and
  P.~Michaleas, ``Spatial temporal analysis of 40,000,000,000,000 internet
  darkspace packets,'' in {\em 2021 IEEE High Performance Extreme Computing
  Conference (HPEC)}, pp.~1--8, 2021.

\bibitem{pisharody2021realizing}
S.~Pisharody, J.~Bernays, V.~Gadepally, M.~Jones, J.~Kepner, C.~Meiners,
  P.~Michaleas, A.~Tse, and D.~Stetson, ``Realizing forward defense in the
  cyber domain,'' in {\em 2021 IEEE High Performance Extreme Computing
  Conference (HPEC)}, pp.~1--7, IEEE, 2021.

\bibitem{huang2018software}
D.~Huang, A.~Chowdhary, and S.~Pisharody, {\em Software-Defined networking and
  security: from theory to practice}.
\newblock CRC Press, 2018.

\bibitem{karvanen2003measuring}
J.~Karvanen and A.~Cichocki, ``Measuring sparseness of noisy signals,'' in {\em
  4th International Symposium on Independent Component Analysis and Blind
  Signal Separation}, pp.~125--130, 2003.

\bibitem{davis18algorithm}
T.~A. Davis, ``Algorithm 1000: Suitesparse:graphblas: Graph algorithms in the
  language of sparse linear algebra,'' {\em ACM Trans. Math. Softw.}, vol.~45,
  Dec. 2019.

\bibitem{soule2004identify}
A.~Soule, A.~Nucci, R.~Cruz, E.~Leonardi, and N.~Taft, ``How to identify and
  estimate the largest traffic matrix elements in a dynamic environment,'' in
  {\em ACM SIGMETRICS Performance Evaluation Review}, vol.~32, pp.~73--84, ACM,
  2004.

\bibitem{zhang2005estimating}
Y.~Zhang, M.~Roughan, C.~Lund, and D.~L. Donoho, ``Estimating point-to-point
  and point-to-multipoint traffic matrices: an information-theoretic
  approach,'' {\em IEEE/ACM Transactions on Networking (TON)}, vol.~13, no.~5,
  pp.~947--960, 2005.

\bibitem{mucha2010community}
P.~J. Mucha, T.~Richardson, K.~Macon, M.~A. Porter, and J.-P. Onnela,
  ``Community structure in time-dependent, multiscale, and multiplex
  networks,'' {\em science}, vol.~328, no.~5980, pp.~876--878, 2010.

\bibitem{tune2013internet}
P.~Tune, M.~Roughan, H.~Haddadi, and O.~Bonaventure, ``Internet traffic
  matrices: A primer,'' {\em Recent Advances in Networking}, vol.~1, pp.~1--56,
  2013.

\bibitem{kepner16mathematical}
J.~{Kepner}, P.~{Aaltonen}, D.~{Bader}, A.~{Bulu{\c{c}}}, F.~{Franchetti},
  J.~{Gilbert}, D.~{Hutchison}, M.~{Kumar}, A.~{Lumsdaine}, H.~{Meyerhenke},
  S.~{McMillan}, C.~{Yang}, J.~D. {Owens}, M.~{Zalewski}, T.~{Mattson}, and
  J.~{Moreira}, ``Mathematical foundations of the graphblas,'' in {\em 2016
  IEEE High Performance Extreme Computing Conference (HPEC)}, pp.~1--9, 2016.

\bibitem{buluc17design}
A.~{Bulu{\c{c}}}, T.~{Mattson}, S.~{McMillan}, J.~{Moreira}, and C.~{Yang},
  ``Design of the graphblas api for c,'' in {\em 2017 IEEE International
  Parallel and Distributed Processing Symposium Workshops (IPDPSW)},
  pp.~643--652, 2017.

\bibitem{fan2004prefix}
J.~Fan, J.~Xu, M.~H. Ammar, and S.~B. Moon, ``Prefix-preserving ip address
  anonymization: measurement-based security evaluation and a new
  cryptography-based scheme,'' {\em Computer Networks}, vol.~46, no.~2,
  pp.~253--272, 2004.

\bibitem{clauset2009power}
A.~Clauset, C.~R. Shalizi, and M.~E. Newman, ``Power-law distributions in
  empirical data,'' {\em SIAM review}, vol.~51, no.~4, pp.~661--703, 2009.

\bibitem{barabasi2016network}
A.-L. Barab{\'a}si {\em et~al.}, {\em Network science}.
\newblock Cambridge university press, 2016.

\bibitem{leland1994self}
W.~E. Leland, M.~S. Taqqu, W.~Willinger, and D.~V. Wilson, ``On the
  self-similar nature of ethernet traffic (extended version),'' {\em IEEE/ACM
  Transactions on Networking (ToN)}, vol.~2, no.~1, pp.~1--15, 1994.

\bibitem{faloutsos1999power}
M.~Faloutsos, P.~Faloutsos, and C.~Faloutsos, ``On power-law relationships of
  the internet topology,'' in {\em ACM SIGCOMM computer communication review},
  vol.~29-4, pp.~251--262, ACM, 1999.

\bibitem{albert1999internet}
R.~Albert, H.~Jeong, and A.-L. Barab{\'a}si, ``Internet: Diameter of the
  world-wide web,'' {\em Nature}, vol.~401, no.~6749, p.~130, 1999.

\bibitem{barabasi1999emergence}
A.-L. Barab{\'a}si and R.~Albert, ``Emergence of scaling in random networks,''
  {\em Science}, vol.~286, no.~5439, pp.~509--512, 1999.

\bibitem{adamic2000power}
L.~A. Adamic and B.~A. Huberman, ``Power-law distribution of the world wide
  web,'' {\em science}, vol.~287, no.~5461, pp.~2115--2115, 2000.

\bibitem{barabasi2009scale}
A.-L. Barab{\'a}si, ``Scale-free networks: a decade and beyond,'' {\em
  science}, vol.~325, no.~5939, pp.~412--413, 2009.

\bibitem{mahanti2013tale}
A.~Mahanti, N.~Carlsson, A.~Mahanti, M.~Arlitt, and C.~Williamson, ``A tale of
  the tails: Power-laws in internet measurements,'' {\em IEEE Network},
  vol.~27, no.~1, pp.~59--64, 2013.

\bibitem{stigler1974studies}
S.~M. Stigler, ``Studies in the history of probability and statistics. xxxiii
  cauchy and the witch of agnesi: An historical note on the cauchy
  distribution,'' {\em Biometrika}, pp.~375--380, 1974.

\bibitem{larson1981urban}
R.~C. Larson and A.~R. Odoni, {\em Urban operations research}.
\newblock No.~Monograph, 1981.

\bibitem{devlin2021hybrid}
P.~Devlin, J.~Kepner, A.~Luo, and E.~Meger, ``Hybrid power-law models of
  network traffic,'' in {\em 2021 IEEE International Parallel and Distributed
  Processing Symposium Workshops (IPDPSW)}, pp.~280--287, IEEE, 2021.

\end{thebibliography}

\appendices
\setcounter{equation}{0}
\renewcommand{\theequation}{\thesection\arabic{equation}}

\end{document}